\documentclass[conference]{IEEEtran}
\IEEEoverridecommandlockouts

\usepackage{cite}
\usepackage{amsmath,amssymb,amsfonts}
\usepackage{graphicx}
\usepackage{textcomp}
\usepackage{xcolor}
\usepackage{multirow}
\usepackage{enumitem}
\usepackage{tabularx}
\usepackage{array}
\usepackage{algorithm}             
\usepackage[noend]{algpseudocode}  
\usepackage{float}                 
\usepackage{upgreek}            
\usepackage{booktabs}
\usepackage{hyperref}

\def\BibTeX{{\rm B\kern-.05em{\sc i\kern-.025em b}\kern-.08em
    T\kern-.1667em\lower.7ex\hbox{E}\kern-.125emX}}

\begin{document}

\title{
  Toward Automated Hypervisor Scenario Generation Based on VM Workload Profiling for Resource-Constrained Environments
}

\author{
  \IEEEauthorblockN{Hyunwoo Kim}
  \IEEEauthorblockA{
    \textit{Intel}\\
    Seoul, South Korea \\
    onion.kim@intel.com
  }
  \and
  \IEEEauthorblockN{Jaeseong Lee}
  \IEEEauthorblockA{
    \textit{Intel}\\
    Seoul, South Korea \\
    jerry.j.lee@intel.com
  }
  \and
  \IEEEauthorblockN{Sunpyo Hong}
  \IEEEauthorblockA{
    \textit{Intel}\\
    Seoul, South Korea \\
    brandon.hong@intel.com
  }
  \and
  \IEEEauthorblockN{Changmin Han}
  \IEEEauthorblockA{
    \textit{Intel}\\
    Seoul, South Korea \\
    duke.han@intel.com
  }
}

\maketitle

\begin{abstract}

  In the automotive industry, the rise of software-defined vehicles (SDVs) has
  driven a shift toward virtualization-based architectures that consolidate
  diverse automotive workloads on a shared hardware platform. To support this
  evolution, chipset vendors provide board support packages (BSPs), hypervisor
  setups, and resource allocation guidelines. However, adapting these static
  configurations to varying system requirements and workloads remain a
  significant challenge for Tier 1 integrators.

  This paper presents an automated scenario generation framework, which helps
  automotive vendors to allocate hardware resources efficiently across multiple
  VMs. By profiling runtime behavior and integrating both theoretical models and
  vendor heuristics, the proposed tool generates optimized hypervisor
  configurations tailored to system constraints.

  We compare two main approaches for modeling target QoS based on profiled data
  and resource allocation: domain-guided parametric modeling and deep
  learning-based modeling. We further describe our optimization strategy using
  the selected QoS model to derive efficient resource allocations. Finally, we
  report on real-world deployments to demonstrate the effectiveness of our
  framework in improving integration efficiency and reducing development time in
  resource-constrained environments.
\end{abstract}


\section{Introduction}

The automotive industry is undergoing a rapid transformation, driven by the
growing demand for intelligent, software-defined vehicles\,
(SDVs)~\cite{liu2022impact,jiacheng2016software,bhatia2019software} that
integrate advanced computing, from real-time decision making to AI-powered
features. This shift has been accompanied by the introduction of
service-oriented software architectures
(SOA)~\cite{papazoglou2007service,valipour2009brief,hustad2021creating}, which
enable modular and dynamic system integration and necessitate the use of
general-purpose electronic control units (ECUs) rather than fixed-function
controllers\cite{liu2022impact}. Traditionally, automotive E/E architectures
have followed a function-oriented approach, where each vehicle function is
handled by a dedicated ECU with tightly integrated
software\cite{mauser2025centralization,bandur2021making}. However, the current
trend is to reduce the number of custom-designed ECUs in favor of
general-purpose SoCs that consolidate multiple functions using
virtualization\cite{kampmann2022optimization}.

To support the increasingly diverse and dynamic software requirements of modern
vehicles, such as real-time operating systems (RTOS), infotainment systems, and
AI workloads, automotive platforms are now relying more heavily on
hypervisor-based virtualization technologies\cite{reinhardt2014embedded}. For
instance, modern cockpit domain controllers commonly deploy separate virtual
machines to isolate real-time control tasks running on RTOS from infotainment
systems, ensuring both safety-critical performance and user experience are
maintained\cite{jiang2024towards,karthik2018hypervisor,li2019acrn}.

Successfully deploying such virtualization-based architectures in vehicles
requires close collaboration across the automotive supply chain, particularly
between chipset vendors and Tier 1
suppliers\cite{zhang2022innovation,volpato2004oem}. The automotive supply chain
is composed of multiple layers, where chipset vendors supply the foundational
hardware platforms (e.g., SoCs), and Tier 1 suppliers integrate these platforms
into domain-specific ECUs or central compute modules for delivery to OEMs.

Chipset vendors, such as Intel, not only provide the hardware but also deliver
essential software packages\cite{trovao2019trends}. These include Board Support
Packages (BSPs), virtualization-ready configurations, and often customized
hypervisors such as ACRN\cite{li2019acrn},
QNX\cite{BlackBerryQNX_HypervisorAutomotive2023}, or
Jailhouse\cite{baryshnikov2016jailhouse}, tailored for automotive use
cases\cite{lozano2023comprehensive}. The vendors typically define recommended
usage patterns for their chipsets. They provide guidance on which operating
systems and workloads should be deployed on virtual machines (VMs), such as
RTOS for control, Linux for
infotainment\cite{macario2009vehicle,jeong2023infotainment}, or AI inference
engines\cite{sohn2024strategy}. They also specify the recommended patterns of
resource allocation based on the size and type of workloads.

\begin{figure}
  \centering
  \includegraphics[width=\linewidth]{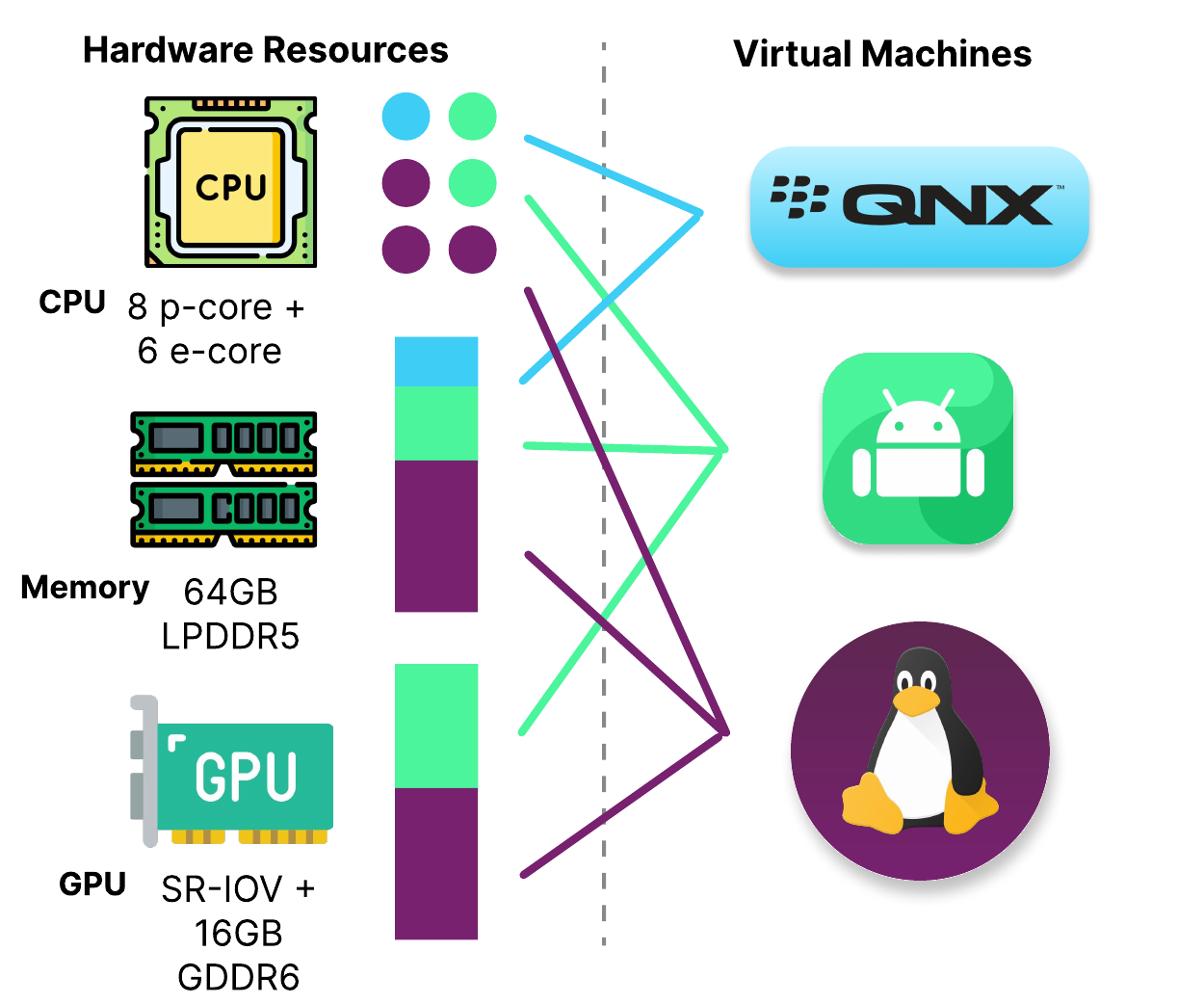}
  \caption{Resource allocation example in automotive hypervisors.}
  \label{fig:resource_allocation}
\end{figure}

Figure~\ref{fig:resource_allocation} illustrates resource allocation in
automotive hypervisors. These allocation recommendations often include detailed
guidance on hypervisor configuration and the allocation of compute, memory, and
I/O resources, including the use of technologies such as Single Root I/O
Virtualization (SR-IOV)\cite{dong2012high,younge2015supporting} to enable
efficient sharing of devices across virtual machines for optimal performance.

However, Tier 1 suppliers often face difficulties in fine-tuning these
allocations, as achieving the \emph{golden ratio} of resource distribution is
non-trivial. Poor allocation can lead to underutilization, resource exhaustion,
or system instability, which are especially problematic in resource-constrained
environments. While chipset vendors may offer guidelines, these are often based
on heuristic or empirical experiments, and adapting them to different vehicle
architectures and workloads remain a significant integration challenge. Vendors
typically provide VM setup patterns specific to their chipset platforms and
supported hypervisors, recommending which operating systems and workloads
should be assigned to each VM. These recommendations often include resource
distribution strategies, such as how to allocate heterogeneous CPU cores (for
example, P-cores and E-cores), partition system memory, and configure I/O
resources using technologies like VirtIO or SR-IOV\cite{li2019acrn}. However,
these configurations are often static and require manual adaptation, making it
difficult for integrators to fully optimize performance across diverse
deployment scenarios.

To address this challenge, we propose an automated scenario generation
framework for hypervisors, designed to assist automotive vendors in allocating
hardware resources efficiently across multiple virtual machines in
resource-constrained environments.

This framework simplifies virtualization integration by offering a
user-friendly interface where vendors can specify target configurations,
including the VM operating system, workload type, peripheral device usage, and
targeting performance or safety requirements. Based on this input, the tool
profiles the runtime behavior of each VM, monitors usage patterns (e.g., CPU,
memory, and I/O loads), and automatically generates an optimized resource
allocation scenario. The generated configuration incorporates both theoretical
guidelines and heuristic knowledge derived from chipset vendor recommendations
and empirical experimentation. While the tool provides an optimized starting
point, vendors retain full control and can fine-tune the final allocation based
on system-specific constraints or design goals. By reducing the need for manual
tuning and abstracting complex resource management processes, this approach
significantly eases the hypervisor integration effort, shortens development
time, and improves system stability and efficiency for automotive vendors.

\medskip
\section{Background}
\label{sec:background}

Optimizing resource allocation in virtualized environments, particularly those
that host dynamic workloads and must meet strict performance or safety
constraints, is a multifaceted challenge. Because most initial hypervisor
configurations are static, they often waste resources, and this inefficiency
becomes critical in embedded and automotive systems where hardware capabilities
are limited.

To address these challenges, recent research has explored both
optimization-based and machine learning-driven approaches to improve the
initial resource allocation for VMs based on workload characteristics.

\smallskip
\subsection{Optimization-Based Approaches}

Optimization has long been studied in computer science, operations research,
and applied mathematics, where the goal is to select the best decision from a
finite or continuous set of alternatives. These methods range from convex
optimization\cite{boyd2004convex,bertsekas2009convex,bubeck2015convex} with
polynomial-time guarantees, through multi-armed bandits
(MAB)\cite{lai1985asymptotically,auer2002finite,vermorel2005multi,galli2023playing,wolsey1999integer}
for online exploration, to mixed-integer linear programming
(MILP)\cite{wolsey1999integer,jain2001algorithms},
metaheuistics\cite{osman1996meta,kirkpatrick1983sa}, and Bayesian optimization
for black-box or combinatorial landscapes. These techniques are widely used for
today's work on resource allocation, scheduling, and machine-learning
hyper-parameter tuning, and they provide the theoretical background in this
paper.

Several works have applied formal optimization techniques to derive
near-optimal static VM configurations. For example, Kampmann et
al.~\cite{kampmann2022optimization} propose a mathematical optimization
framework for resource allocation in an automotive service-oriented
architecture. Their solution assigns CPU cores and scheduling slots to
automotive software services, balancing competing objectives such as minimizing
power consumption and reducing worst-case end-to-end latency.

Other studies explore more general-purpose optimization frameworks. Sun et
al.~\cite{sun2020optimal} employ convex optimization and Lyapunov stability
theory to derive stable and efficient VM placement strategies in enterprise
environments. Similarly, Dubey et al.~\cite{dubey2023resource} utilize an
enhanced Genetic Algorithm (GA) in CloudSim to optimize for both energy
consumption and job makespan, demonstrating the efficacy of evolutionary
techniques in finding balanced allocation solutions.

Kabir et al.~\cite{kabir2023virtualization} propose a virtual prototyping
framework for automotive use cases, leveraging digital twins and virtual ECUs
to simulate different VM resource allocations. Their system enables early-stage
profiling and optimization of resource distribution within a controlled
simulation environment, helping integrators toward effective static
configurations before actual hardware deployment.

\smallskip
\subsection{Machine Learning for Resource Prediction}

Machine learning (ML) techniques are increasingly used to model and predict VM
resource requirements based on historical or runtime profiling data.

Rao et al.~\cite{rao2009vconf} present a model-based reinforcement-learning
(RL) agent that automatically tunes each VM's CPU share, vCPU count, and memory
quota on Xen hosts. Building on this research, numerous RL-based hypervisor
resource-allocation techniques have since been proposed
\cite{hummaida2022scalable,ma2022real,shah2020multiagent}.

Khan et al.~\cite{khan2022workload} and Gong et al.\cite{gong2024dynamic}
extend this idea by incorporating deep learning and reinforcement learning to
dynamically infer workload characteristics and guide resource management
policies. These models analyze runtime metrics such as CPU utilization, memory
footprint, and I/O load to anticipate resource bottlenecks and adapt VM
allocations accordingly.

Vhatkar et al.~\cite{vhatkar2024improved} introduce an end-to-end AI-driven
cloud management framework, integrating predictive deep learning models with
recycling mechanisms to continuously refine VM placement and utilization.
Though primarily cloud-oriented, this layered architecture exemplifies the
potential of AI in automating complex resource management decisions, even in
embedded or automotive contexts.

\smallskip
\subsection{Toward Scenario-Aware Static Allocation}

Together, these studies illustrate a growing interest in using empirical
profiling, optimization theory, and machine learning to reduce the manual
burden of hypervisor configuration. In particular, the ability to generate or
recommend static VM scenarios based on workload profiling is increasingly seen
as a necessary step toward scalable and efficient virtualization-based
architectures, especially for hardware-constraint domain such as automotive
industry.

\medskip
\section{Problem Statement}
\label{sec:problem_statement}

\smallskip
\subsection{Challenges in Manual Resource Allocation}

Virtualization in automotive systems introduces complex resource management
challenges, especially when distributing limited hardware resources among
multiple VMs with heterogeneous workloads. In practice, manual allocation often
leads to suboptimal system performance and reliability. This section
categorizes the common pitfalls observed in VM resource allocation.

\smallskip
\subsubsection{vCPU Misallocation}

Modern automotive SoCs often feature heterogeneous
cores\cite{nikov2015evaluation,burgio2017software}, such as performance (P)
cores and efficiency (E) cores. Assigning latency-sensitive workloads, such as
real-time operating systems (RTOS), to E-cores can lead to missed deadlines and
performance degradation. Conversely, over-allocating P-cores to non-critical
domains may lead to inefficient use of compute resources. Fine-grained control
over core assignment is essential but challenging to manage manually.

\smallskip
\subsubsection{Memory Misallocation}

Incorrect memory sizing is another frequent issue. VMs with insufficient memory
may trigger excessive swapping or paging, degrading application responsiveness
and system stability\cite{wang2023efficient,sawamura2015evaluating}. In
particular, assigning too little memory to the primary domain (e.g., Dom0 or
Service VM) can offload graphics or device management tasks to shared memory,
increasing latency and CPU load.

\smallskip
\subsubsection{SR-IOV Misconfiguration}

Even with vendor-approved SR-IOV configurations, subtle parameter mistakes can
cripple performance. A common pitfall is allocating insufficient memory to the
physical function (PF). For instance, some users might allocate less than 256
MB of PF memory, even when the GPU vendor recommends a minimum of 1\,GB. With
the PF starved of on-chip memory, command buffers from its virtual functions
(VFs) must spill into system RAM, forcing the blitter engine to perform extra
DMA hops. The resulting queue thrashing manifests as dropped frames, sporadic
stutters, and, in severe cases, temporary device loss until the scheduler
recovers.

\medskip
\section{Automated Allocation Proposal}

To solve the problem outlined in Section~\ref{sec:problem_statement}, we design
an end-to-end \emph{automated resource-allocation pipeline} that moves from raw
hardware data to a validated, ready-to-deploy hypervisor scenario. The workflow
proceeds through five stages:

\begin{enumerate}[leftmargin=*,label=\textbf{\arabic*.}]
  \item \textbf{Hardware inspection.}\label{step:hardware_inspection}
        The operator provides the XML board configuration file, which the framework
        parses to reconstruct the CPU topology, physical-memory limits, GPU capabilities,
        and the peripheral inventory of the target platform.

  \item \textbf{High-level scenario configuration.}
        Building on the board config from Step~\ref{step:hardware_inspection}, the operator declares a high level scenario
        configurations for each VM, such as guest OS, workload class, target QoS metrics, and peripheral requirements.
        This generates scripts to profile the VM's resource usage.

  \item \textbf{VM resource usage profiling.}
        Each VM then boots with generous provisional resources and an instrumented
        script records time-series traces of resources(CPU utilization, working-set size, and GPU load).
        These traces are condensed into concise workload profiles.

  \item \textbf{Resource-allocation optimization and scenario generation.}
        An optimization engine combines the workload profiles with the discovered
        hardware constraints to compute an optimal resource allocation value, such as CPU cores, memory capacity,
        and GPU slices, that satisfies the declared QoS targets and maximizes overall utilization efficiency.
        A launch script is then generated with the hypervisor configuration tool based on the computed allocation.

  \item \textbf{Launch VMs and refinement.}
        Launch the VMs with the generated launch script,
        The operator can perform iterative heuristic refinements before deploying the VMs in production.
\end{enumerate}

\begin{figure*}
  \centering
  \includegraphics[width=0.85\textwidth]{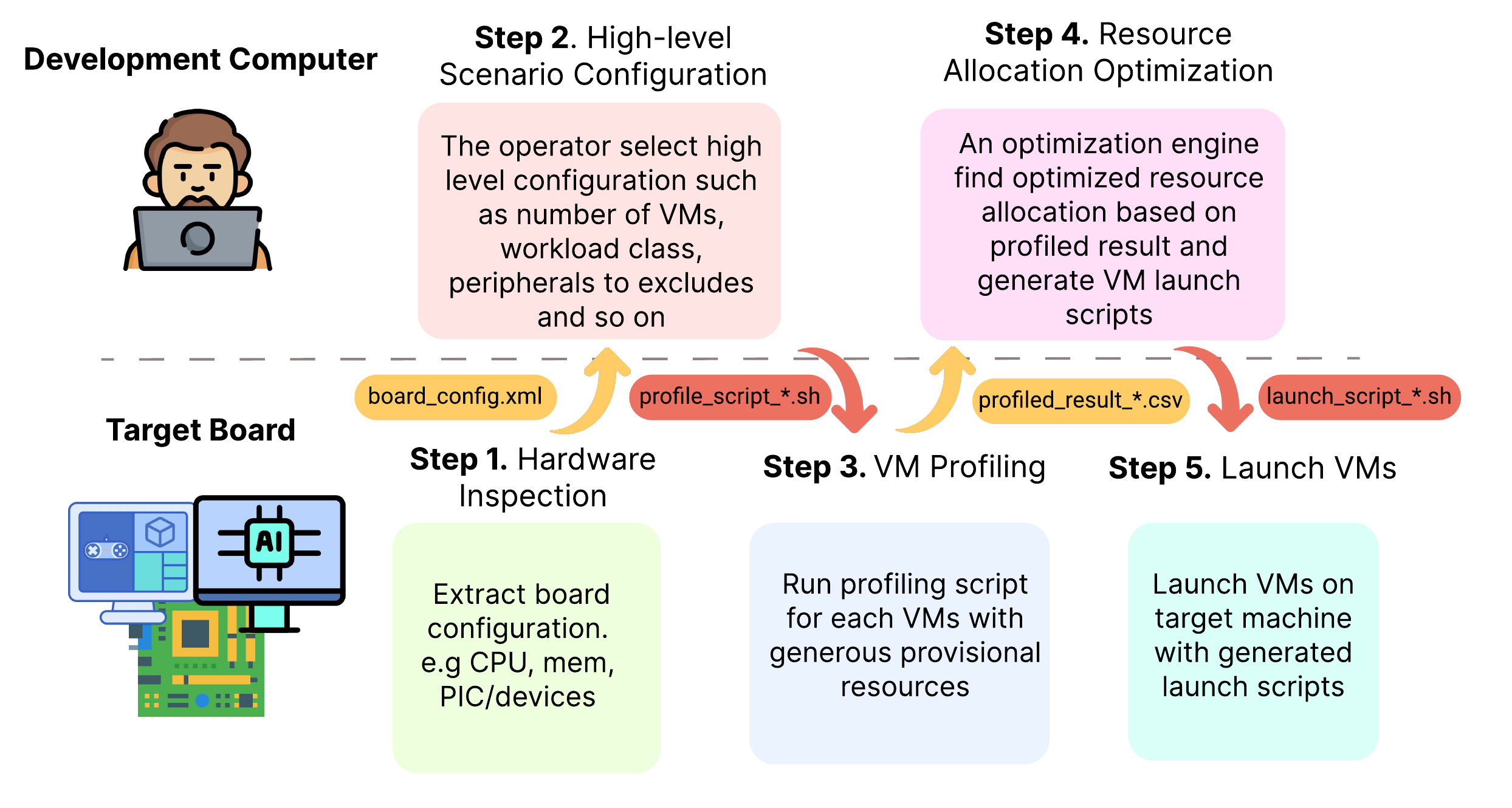}
  \caption{Automated hypervisor scenario generation pipeline with optimized resource allocation.}
  \label{fig:optimization_process}
\end{figure*}

These five stages are summarized in Figure~\ref{fig:optimization_process}. This
pipeline eliminates manual trial and error, shortens provisioning time, and
delivers allocations that are both performance aware and resource efficient.

\medskip
\section{Resource Allocation Proposal}
\label{sec:resource_allocation_proposal}

This section proposes a modeling and optimization framework that builds on the
section~\ref{sec:problem_statement}. The design clarifies \emph{what we
  optimize}, \emph{how we measure it}, and \emph{how profiled workload behaviour
  is converted into a predictive QoS-driven objective function} used for
constrained resource allocation across multiple VMs.

\smallskip
\subsection{Design Challenges}

Automating VM resource allocation demands a framework that can \emph{observe,
  infer, and act}: it must profile each workload, predict how candidate
allocations will impact QoS, and then recommend an efficient split of
resources. Four research challenges must be addressed:

\begin{itemize}[leftmargin=*]
  \item \textbf{Instrumentation scope.} Which signals should the profiler capture? Examples include CPU utilization, working-set size, I/O bandwidth, GPU occupancy, latency distributions, or a carefully weighted subset of these.
  \item \textbf{optimization objectives.} Should the allocator minimize latency, maximize throughput, balance resource utilization, or trade off several QoS metrics according to workload priorities?
  \item \textbf{Predictive modeling.} How can raw traces and candidate allocations be translated into accurate forecasts of the target metrics? The choice of modeling technique is also crucial, whether it be analytical equations, statistical regressors, or learned models, such as neural networks or decision trees, depending on the desired trade-off between accuracy and complexity.
  \item \textbf{Search strategy.} Given the model, which algorithm most effectively navigates the mixed discrete-continuous search space to find near-optimal allocations within practical time constraint? Possible choices include greedy heuristics, MILP, backtracking, and reinforcement learning.
\end{itemize}

Solving these challenges is critical to delivering a robust, end-to-end
allocator that scales across heterogeneous hardware and diverse workload
profiles.

\smallskip
\subsection{Design Goals}
\begin{enumerate}
  \item \textbf{QoS-Aware Allocation}: Allocate CPU cores (including P-/E-core heterogeneity), memory, and GPU SR-IOV slices so that each VM meets (or approaches) workload-specific QoS targets.
  \item \textbf{Interpretable Modeling}: Use a modeling approach that is simple enough to explain, tune, and justify in academic and industrial settings.
  \item \textbf{Data Efficiency.} Produce usable QoS response models from \emph{limited profiling data}, as is often the case in embedded, automotive, or edge deployments where exhaustive sweeps are expensive.
  \item \textbf{Optimization Under Hard Constraints}: Respect global resource budgets (CPU cores by type, total memory, GPU capacity/instances) and any indivisibility or affinity constraints (e.g., core pinning, NUMA domains, GPU function counts).
\end{enumerate}

\smallskip
\subsection{Notation}
Table~\ref{tab:notation} lists the symbols and abbreviations that will be used
throughout the remainder of this paper.

\begin{table}[htbp]
  \caption{Summary of notation.}
  \label{tab:notation}
  \centering
  \begin{tabularx}{\linewidth}{|>{\centering\arraybackslash}m{1.2cm}|X|}
    \hline
    \textbf{Symbol}                              & \textbf{Meaning}                                                                                                                                                                       \\
    \hline
    $N$                                          & Number of VMs; index $i = 1,\dots,N$.                                                                                                                                                  \\
    $\mathbf{r}_i$                               & Resource allocation vector for VM$_i$, e.g., $\mathbf{r}_i = (c_i^P, c_i^E, m_i, g_i)$ where $c_i^P$/$c_i^E$ are P-/E-cores, $m_i$ is memory (GiB), and $g_i$ is the GPU SR-IOV share. \\
    $\mathbf{p}_i$                               & Profiled workload vector for VM$_{i}$                                                                                                                                                  \\
    $K_i$                                        & Number of QoS metrics tracked for VM$_{i}$; index $k = 1,\dots,K_i$.                                                                                                                   \\
    $Q_{i,k}$                                    & \emph{Measured} QoS value (e.g., latency, FPS) for VM$_{i}$, metric~$k$, under allocation $\mathbf{r}_i$.                                                                              \\
    $\widehat{Q}_{k}(\mathbf{r}_i,\mathbf{p}_i)$ & \emph{Predicted} QoS for metric~$k$ of VM$_i$, computed by the response model based on the allocated resources $\mathbf{r}_i$ and the profiled resource usage $\mathbf{p}_i$           \\
    $S_{k}$                                      & normalized QoS score in $[0,1]$ for metric~$k$ (higher~$\Rightarrow$ better).                                                                                                          \\
    $w_{i,k}$                                    & Importance weight for QoS metric~$k$ of VM$_{i}$ (weights may sum to~1 within each VM).                                                                                                \\
    $U_{i,k}$                                    & \emph{Measured} utilization value (e.g., cpu, mem) in percentage under allocation $\mathbf{r}_i$.                                                                                      \\
    $\widehat{U}_{k}(\mathbf{r}_i,\mathbf{p}_i)$ & Predicted utilization for metric~$k$ of VM$_i$, computed by the response model based on the allocated resources $\mathbf{r}_i$ and the profiled resource usage $\mathbf{p}_i$          \\
    $v_{i,k}$                                    & Importance weight for utilization metric~$k$ of VM$_{i}$ (weights may sum to~1 within each VM).                                                                                        \\
    \hline
  \end{tabularx}
\end{table}

\smallskip
\subsection{Composite Objective}

The goal is to maximize the aggregate utility across all \(N\) VMs:

\begin{equation}
  \label{eq:objective}
  \begin{aligned}
    \max_{\{\mathbf{r}_i\}_{i=1}^{N}} \quad & \sum_{i=1}^{N} \mathrm{VMOptScore}_{i}(\mathbf{r}_{i}) \\
  \end{aligned}
\end{equation}

\noindent
For each VM the optimization score has two parts:
(1)~a \emph{performance} term that rewards meeting QoS targets, and
(2)~a \emph{utilization} term that encourages efficient use of the allocated
resources:

\begin{equation}
  \label{eq:vm_opt_score}
  \begin{split}
    \mathrm{VMOptScore}_{i}(\mathbf{r}_{i}) =
     & \underbrace{\mathrm{VMPerfScore}_{i}(\mathbf{r}_{i})}_{\text{QoS utility}}                                  \\
     & + \lambda_{\text{util}}\, \underbrace{\mathrm{VMUtilScore}_{i}(\mathbf{r}_{i})}_{\text{efficiency utility}}
  \end{split}
\end{equation}

where \(\lambda_{\text{util}}\) balances performance against efficiency.

\smallskip
\noindent
The performance utility is a weighted sum of normalized QoS scores:

\[
  \mathrm{VMPerfScore}_{i}(\mathbf{r}_{i}) =
  \sum_{k=1}^{K_i}
  w_{i,k}\,
  S_{k}\!\bigl(\widehat{Q}_{k}(\mathbf{r}_{i},\mathbf{p}_{i})\bigr),
\]

\noindent
where \(w_{i,k}\) captures the importance of the \(k\)-th QoS metric for
VM$_{i}$.

For each VM the allocation vector is
\(\mathbf{r}_{i}=(c_i^{P},\,c_i^{E},\,m_i,\,g_i,...)\), with
\begin{itemize}
  \item \(c_i^{P}\): number of assigned \textbf{P-cores};
  \item \(c_i^{E}\): number of assigned \textbf{E-cores};
  \item \(m_i\): amount of allocated \textbf{memory};
  \item \(g_i\): share of \textbf{GPU SR-IOV} resources.
\end{itemize}

The allocations must satisfy the capacity limits and VM-specific feasibility
rules:

\[
  \text{s.t.}\;
  \left
  \{
  \begin{alignedat}{2}
    \sum_{i=1}^{N} c_i^{P} & \le C^{P}_{\text{tot}}, \qquad &
    \sum_{i=1}^{N} c_i^{E} & \le C^{E}_{\text{tot}},                                        \\[2pt]
    \sum_{i=1}^{N} m_i     & \le M_{\text{tot}},            &
    \sum_{i=1}^{N} g_i     & \le G_{\text{tot}},                                            \\[2pt]
    \mathbf{r}_i           & \in \mathcal{F}_i              &   & \forall i\in\{1,\dots,N\}
  \end{alignedat}
  \right.
\]

where \(\mathcal{F}_{i}\) encodes VM-specific feasibility constraints (e.g.,
minimum/maximum core counts, NUMA pinning rules, GPU slice granularity).

For example, a latency-sensitive API VM could use
\[
  \mathrm{VMPerfScore}_{\text{web VM}}(\mathbf{r}_i)=
  0.4\,S_{\text{lat}}\!\bigl(\widehat{L}_i(\mathbf{r}_i)\bigr)+
  0.1\,S_{\text{thr}}\!\bigl(\widehat{T}_i(\mathbf{r}_i)\bigr)+\dots
\]
where \(\widehat{L}_i\) is the predicted \(p_{99}\) latency,
\(S_{\text{lat}}(L)=\max(0,1-L/L_{\text{SLO}})\), and the weights (here
\(0.4,0.1,\dots\)) sum to one to reflect metric priorities.

\smallskip
\noindent
The efficiency term is formed in an analogous way, by combining the predicted
utilization of each resource with weights that reflect its importance to the
workload:

\[
  \mathrm{VMUtilScore}_{i}(\mathbf{r}_{i}) =
  \sum_{k=1}^{K_i}
  v_{i,k}\,
  \widehat{U}_{i,k}\!\bigl(\mathbf{r}_{i},\mathbf{p}_{i}\bigr),
\]

\noindent
where \(\widehat{U}_{i,k}\in[0,1]\) is the predicted fraction of resource
\(k\) actually consumed under allocation \(\mathbf{r}_{i}\), and the weights
\(v_{i,k}\) (with \(\sum_k v_{i,k}=1\)) capture how much the workload cares
about utilising each resource efficiently.

As a simple example, suppose profiling shows that the workload peaks at 2\,GiB
of memory; if the allocator grants 4\,GiB, the memory-utilization score is
\(\widehat{U}_{\text{mem}} = 2/4 = 0.5\). If memory efficiency is deemed more
important than CPU efficiency, one might choose \(v_{\text{mem}} = 0.5\) and
\(v_{\text{cpu}} = 0.1\), giving

\[
  \mathrm{VMUtilScore}_{i} =
  0.5 \times \widehat{U}_{\text{mem}}
  \;+\;
  0.1 \times \widehat{U}_{\text{cpu}}
  \;+\;\dots,
\]

\noindent
so the score increases when the VM uses its allotted memory and CPU more
proportionally to their respective priorities.

\smallskip
\subsection{QoS parameters by Workload Class}

Because each VM serves a different purpose, web serving, gaming, AI inference,
and so on, the relevant QoS metrics and their weights \(w_{i,k}\) must be
chosen accordingly. Table~\ref{tab:qos_parameters} lists typical primary and
secondary metrics for common workload classes, along with brief guidance on how
their scores are usually composed. These templates can be taken as a starting
point and tweaked to match specific customer SLOs.

\begin{table*}[htbp]
  \caption{QoS dimensions and scoring guidance by workload class.}
  \label{tab:qos_parameters}
  \centering
  \begin{tabularx}{\textwidth}{|l|l|l|X|}
    \hline
    \textbf{Workload Class} & \textbf{Primary QoS}   & \textbf{Secondary QoS}       & \textbf{Scoring Notes}                                                                  \\
    \hline
    Gaming                  & FPS; frame-time jitter & Input latency, VRAM pressure & Strongly weight FPS; penalize stutter (std.~dev.~of frame time).                        \\
    AI Inference            & Latency (p50/p95/p99)  & Throughput (req/s, tokens/s) & Latency weight dominates when SLO tight; throughput weight grows when batching allowed. \\
    Web / Microservices     & Tail latency (p95/p99) & RPS under SLO                & Classic service SLO; can reuse Apdex-style scoring.                                     \\
    RTOS / Control          & Deadline miss ratio    & Jitter                       & Hard deadline; binary or steep penalty beyond threshold.                                \\
    \hline
  \end{tabularx}
\end{table*}

\smallskip
\subsection{Profiling Data Collection}
During the \emph{Profiling Phase}, each workload is executed in isolation with
ample resources to characterize sensitivity curves. Key signals to capture:
\begin{itemize}
  \item \textbf{CPU}: average/peak utilization; scaling when cores removed; P-/E-core sensitivity.
  \item \textbf{Memory}: working-set size (WSS), peak resident-set size(RSS), swap incidence, page-fault rate.
  \item \textbf{GPU}: utilization by engine (compute, copy, graphics), VRAM footprint, PCIe bandwidth.
\end{itemize}
Profiling sweeps should vary one or more resource dimensions to reveal breakpoints (e.g., memory~$<\text{WSS}$ triggers paging; GPU slice~$<\text{fit}$ triggers batching delay).

\smallskip
\subsection{Optimization}
With the QoS and utilization models defined in
Section~\ref{sec:resource_allocation_proposal},\footnote{Recall
  $\mathrm{VMOptScore}_{i}(\mathbf{r}_{i})$ combines normalized QoS utility and
  utilization efficiency.} the remaining task is to choose an allocation
$\{\mathbf{r}_{i}^{\star}\}_{i=1}^{N}$ that maximizes the global objective
while respecting the hard resource constraints.

\smallskip
\paragraph{Why backtracking?}

To solve this optimization problem, we adopt a depth-first \textbf{backtracking
  search} with pruning. This approach is chosen for four key reasons:

\begin{itemize}
  \item \textbf{Heterogeneous quanta.}  Discretising continuous resources
        aligns naturally with the incremental assignment explored by
        backtracking.
  \item \textbf{Non-convex score surface.}  Empirically, QoS curves are
        \emph{non-monotonic} and \emph{non-convex}; classical convex solvers
        therefore provide no optimality guarantees.
  \item \textbf{Feasibility-aware pruning.}  Unlike exhaustive enumeration or
        naive dynamic programming, backtracking rejects partial allocations
        that already violate resource budgets, reducing the search tree
        substantially.
  \item \textbf{Implementation simplicity.}  The algorithm requires only
        (i)~a priority order for resources, and (ii)~an admissible upper bound
        on the best obtainable score from any partial assignment; both are
        easy to express in <100 lines of Python.
\end{itemize}

\smallskip
\paragraph{Alternative heuristics and their limits}
\begin{itemize}
  \item \emph{Greedy} rounding is fast but may be trapped in poor local
        optima.
  \item \emph{Branch-and-bound} (BnB) can dominate backtracking if a tight
        bound can be derived—something elusive for our non-convex QoS models.
  \item \emph{Meta-heuristics} (e.g., simulated annealing, genetic algorithms)
        remain future work; early tests showed slower convergence than
        backtracking under realistic VM counts ($N\!\le\!16$).
\end{itemize}

Overall, depth-first backtracking with feasibility pruning offered the best
trade-off between optimality gap and engineering effort for our prototype.

\medskip
\section{QoS modeling with Profiled-Metric Datasets}
\label{sec:qos_modeling_with_profiled_metric_datasets}

This section describes how our proof-of-concept dataset was collected and how
it underpins the workload-specific QoS response models that drive the optimizer
from Section~\ref{sec:resource_allocation_proposal}.

\smallskip
\subsection{Workload Coverage}
We focus on the \emph{three} software classes most prevalent in modern
automotive stacks, especially for HMI and infotainment:

\begin{enumerate}
  \item \textbf{Gaming}:
        key QoS = FPS.  Multiple Unity and Unreal titles
        were profiled under varying CPU, memory, and GPU-slice allocations.
  \item \textbf{AI Inference}:
        key QoS = token throughput \((\text{tokens}\,\mathrm{s}^{-1})\).
        We benchmarked LLMs from 1.5B to 8B parameters across 3
        quantisation levels (16, 8, 4 bit).
  \item \textbf{Web / MSA API}:
        key QoS = tail-latency (p99) per request.
        A micro-service bundle—NGINX, PostgreSQL, and a Flask API—was driven by
        replayed traces at 5k req/s.
\end{enumerate}

\begin{figure*}[htbp]
  \centering
  \includegraphics[width=\textwidth]{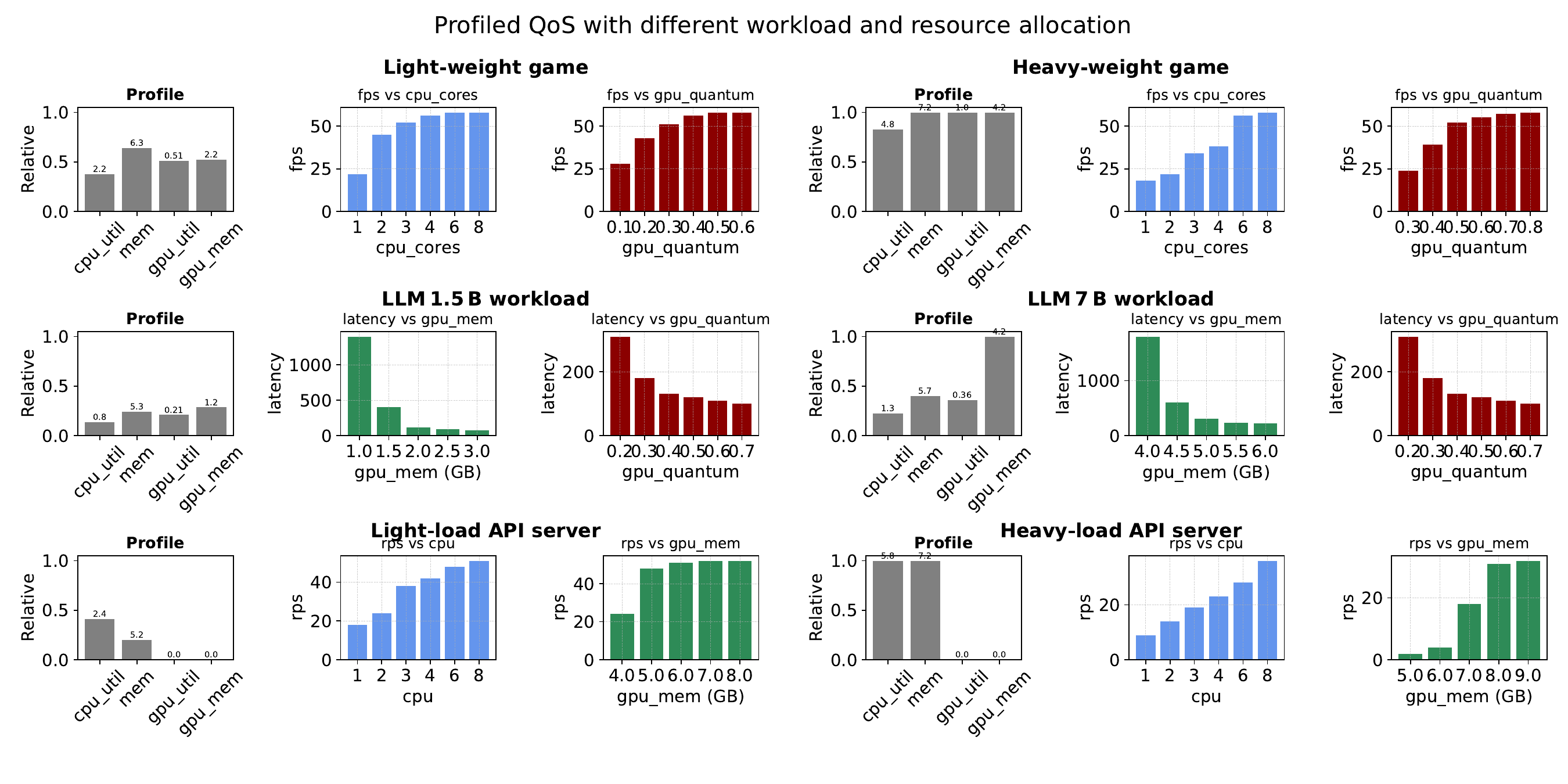}
  \caption{Dataset of profiled workloads across multiple application types, annotated with their respective target~QoS levels.}
  \label{fig:monitor_grid}
\end{figure*}

\smallskip
\subsection{Dataset Collection}
For the data to generalize across deployments, each workload class must satisfy

\begin{itemize}
  \item \emph{Diverse workload sizes}: different games, model scales, and API
        mixes to expose scaling trends;
  \item \emph{Resource sweeps}: systematic variation of cores, memory, and GPU
        time-slices to reveal bottlenecks and tipping points;
  \item \emph{Balance \& uniqueness}: no duplicated rows; uniform sampling of
        the feasible region to avoid skewed regressors.
\end{itemize}

We collect a dataset of profiled metrics for different workload categories and
sizes at each target-QoS level. Figure \ref{fig:monitor_grid} visualizes the
monitored QoS metrics, organized by application type and workload magnitude.

\smallskip
\subsubsection{Collection Procedure}
Each run captures

\[
  \bigl\langle \text{workload ID},\;
  \mathbf{r},\;
  \mathbf{p},\;
  \underbrace{Q}_{\text{QoS}},\;
  \underbrace{\mathbf{u}}_{\text{fine-grained utilization}}
  \bigr\rangle,
\]
where \(\mathbf{r}\) is the allocation vector, \(\mathbf{p}\) the profiled
features (e.g., working-set size, layer count), and \(\mathbf{u}\) resource
utilization snapshots. In total we gathered \(3\times100\approx300\) unique
observations, which is enough to train lightweight regressors for validation.

\smallskip
\subsubsection{Example Rows}
Table~\ref{tab:dataset_example} illustrates the schema with one row per
workload class (truncated for brevity).

For the proof of concept, we evaluated QoS of well-used software use-cases for
automotive system, which is Gaming, AI-inference, and API server with MSA
architecture. For Gaming, we measured the FPS as our prime QoS and collected
the FPS depending on different resource allocation. For AI inference, we tested
with the different size of LLM models, from 1.5b to 8b with different
quantization techniques, measured the throughput(tokens/sec) as a main QoS. For
API service, multiple container service (DB, API server and so on), are loaded
on system. the latency of the api request was measured as key QoS.

\begin{table*}[htbp]
  \caption{Dataset example (one representative row per class; full set has
    100 rows each).}
  \label{tab:dataset_example}
  \centering
  \begin{tabularx}{\textwidth}{|l|X|X|l|}
    \hline
    \textbf{Class}                               &
    \textbf{Profiled Metrics ($\mathbf{p}$)}     &
    \textbf{Allocation ($\mathbf{r}$)}           &
    \textbf{QoS}                                   \\
    \hline
    Web API                                      &
    mem\_rss = 5\,GiB; swap = 0; CPU (P/E) = $\{90,95,10,8\}\,\%$;
    GPU util = $\{20,10,0\}\%$; GPU-mem = 2\,GiB &
    mem = 8\,GiB; $c^{P}=4$, $c^{E}=2$;
    GPU slice = 30\,\%; GPU-mem = 2\,GiB         &
    250\, $\upmu$s (p99 latency)                   \\[2pt]
    Gaming                                       &
    CPU frame time p95 = 12 ms; VRAM = 1.6\,GiB;
    GPU busy = 78\,\%                            &
    mem = 6\,GiB; $c^{P}=6$, $c^{E}=0$;
    GPU slice = 40\,\%                           &
    58 FPS                                         \\[2pt]
    AI Inf.                                      &
    Model = Llama-2-7B Q4;
    ctx len = 2048; peak mem = 9\,GiB            &
    mem = 12\,GiB; $c^{P}=8$, $c^{E}=4$;
    GPU slice = 60\,\%                           &
    102 tokens/s                                   \\
    \hline
  \end{tabularx}
\end{table*}

\smallskip
\subsection{QoS modeling}

We seek functions $\widehat{Q}{k}(\mathbf{r}_i,\mathbf{p}_i)$ that predict QoS
from an allocation vector. Two model families are considered.

\begin{enumerate}[label=\arabic*.]
  \item \textbf{Domain-Guided Parametric / Regression Models (Chosen Baseline)}\\
        Interpretability, low data requirement, monotonicity enforcement. Example forms:
        \begin{itemize}
          \item Saturation: $Q = Q{\max}\bigl(1 - e^{-\alpha x}\bigr)$.
          \item Memory knee: $\text{Latency}(m) = L_0 + L_{\text{page}},\exp\bigl(\beta \max(0,
                  (W-m)/W)\bigr)$\cite{shin2024beyond}.
          \item CPU scaling: $\text{Throughput}(c) = {T_{\infty} c}/{(k+c)}$
                (Michaelis--Menten)\cite{shin2024beyond}.
        \end{itemize}
  \item \textbf{Deep Learning Models}\cite{sodhro2019artificial,mao2020ai} (e.g., MLPs, NAMs, sequence models, mixture-density nets).
\end{enumerate}

\smallskip
\subsubsection{First Approach: Domain-Guided Parametric QoS Models}
\label{sec:domain_parametric_qos}

\smallskip
\paragraph{Empirical regularities}
\begin{itemize}
  \item \textbf{Under-provisioning} ($r<r_{\text{prof}}$):
        QoS improves \emph{approximately linearly} with each extra unit of the
        bottleneck resource.
  \item \textbf{Over-provisioning} ($r>r_{\text{prof}}$):
        returns \emph{diminish}; QoS approaches the profiled value asymptotically.
\end{itemize}

These patterns hold for memory, CPU cores, GPU SR-IOV quanta, network
bandwidth, \dots. We therefore model each resource $k\!\in\!\mathcal{R}$ with a
\emph{resource-specific impact factor} $F_k(r_k)\in(0,\,1]$ that scales the
profiled QoS.

\smallskip
\paragraph{Generic impact factor}

Let \( r_{\text{min},k}\, (< r_{\text{prof},k}) \) be the lowest usable
allocation (e.g., swap threshold, one CPU core, one GPU slice).

For an allocation $r_k$ of resource~$k$ we set
\[
  F_k(r_k)=
  \begin{cases}
    \displaystyle
    \alpha_k\bigl(r_k-r_{\text{min},k}\bigr),              & r_k < r_{\text{prof},k},   \\[10pt]
    1 - c_k\exp\!\bigl[d_k\,(r_k-r_{\text{prof},k})\bigr], & r_k \ge r_{\text{prof},k},
  \end{cases}
\]
where
\[
  c_k \;=\; 1-\alpha_k\,\bigl(r_{\text{prof},k}-r_{\text{min},k}\bigr),\qquad
  d_k \;=\; -\frac{\alpha_k}{c_k},
\]
so that $F_k$ is \emph{continuous} and \emph{differentiable} at
$r_k=r_{\text{prof},k}$ and monotone w.r.t.\ $r_k$.

\begin{figure}[tb]
  \centering
  \includegraphics[width=\linewidth]{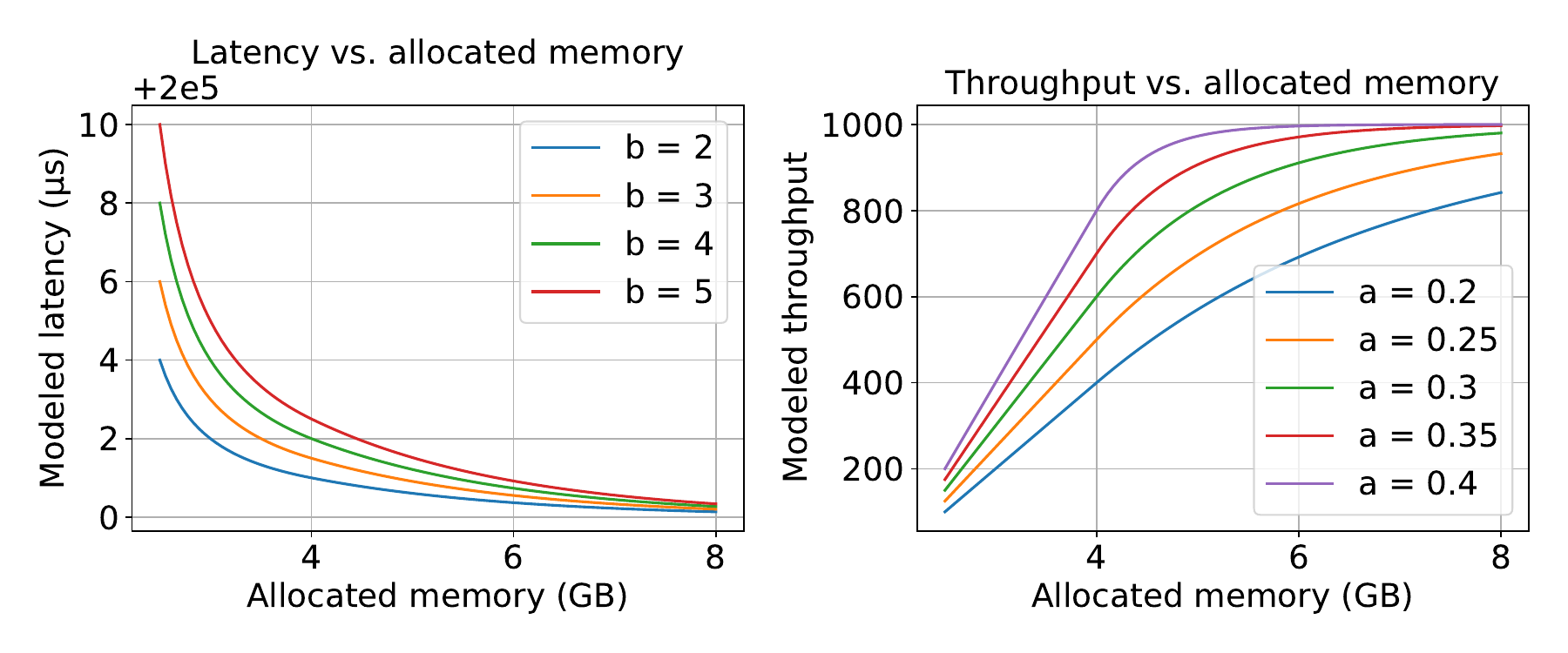}
  \caption{Impact of the tuning parameter~$\alpha_k$ on the QoS curve for memory allocation.  A higher $\alpha_k$ steepens the initial decline while preserving the overall curve shape.}
  \label{fig:qos_trend}
\end{figure}

\smallskip
\emph{Tuning knob.} The parameter $\alpha_k\!>\!0$ governs the slope of the QoS curve
in the under-provisioned region: the larger $\alpha_k$ is, the faster QoS deteriorates
as the allocation approaches $r_{\text{min},k}$. Figure \ref{fig:qos_trend} depicts this
effect for memory allocation under several $\alpha_k$ settings.

\smallskip
\paragraph{Composing multiple resources}

We assume separability and compute an overall multiplicative factor
\begin{equation}
  \label{eq:qos_impact_factor}
  F(\mathbf{r}) \;=\; \prod_{k\in\mathcal{R}} F_k(r_k).
\end{equation}

\begin{itemize}
  \item \textbf{Throughput-type QoS (higher-is-better)}
        \[
          T(\mathbf{r}) = T_{\text{prof}} \times F(\mathbf{r}).
        \]
  \item \textbf{Latency-type QoS (lower-is-better)}
        \[
          L(\mathbf{r}) = \frac{L_{\text{prof}}}{F(\mathbf{r})}.
        \]
\end{itemize}

Both expressions recover the profiled measurement when
$\mathbf{r}=\mathbf{r}_{\text{prof}}$ and respect the observed
under-/over-provision behaviour across heterogeneous resources.

\smallskip
\paragraph{Calibration}

With one extra measurement below $r_{\text{prof},k}$, or a least-squares fit
over many samples, we solve for each $\alpha_k$. All parameters keep clear
physical meaning—“penalty per unit under-allocation”—making the model
interpretable and easy to extend (e.g., by turning the exponential exponent
into an additional parameter if future data warrant).

\smallskip
\subsubsection{Second Approach: Deep-Learning Baseline}

To gauge how far a purely data-driven method can push predictive accuracy, we
also train a \emph{multi-layer perceptron} (MLP) that maps the concatenated
feature vector \((\mathbf{r},\mathbf{p})\) to a QoS estimate
\(\widehat{Q}=f_{\boldsymbol{\theta}}(\mathbf{r},\mathbf{p})\):
\[
  (\mathbf{r},\mathbf{p})
  \;\;\xrightarrow{\;\;\text{MLP}(d_{\text{in}}\!\to\!h_1\!\to\!h_2\!\to\!1)\;\;}\;\;
  \widehat{Q},
\]
where \(d_{\text{in}}=|\mathbf{r}|+|\mathbf{p}|\) and \(h_1,h_2\) are hidden
layer widths (ReLU activations, batch-norm, and dropout applied). Model
parameters \(\boldsymbol{\theta}\) are learned by minimizing the mean-squared
error (MSE) over the profiled dataset~\(\mathcal{D}\):
\[
  \min_{\boldsymbol{\theta}}
  \; \frac{1}{|\mathcal{D}|}
  \sum_{(\mathbf{r},\mathbf{p},Q)\in\mathcal{D}}
  \!\!\bigl(Q-f_{\boldsymbol{\theta}}(\mathbf{r},\mathbf{p})\bigr)^{2},
\]
optimized with Adam (\(\eta=10^{-3}\), early stopping on a validation split).

\smallskip
\subsubsection{Model Comparison and Our Choice}
\label{sec:model_comparison}

\begin{figure}[ht]
  \centering
  \includegraphics[width=\linewidth]{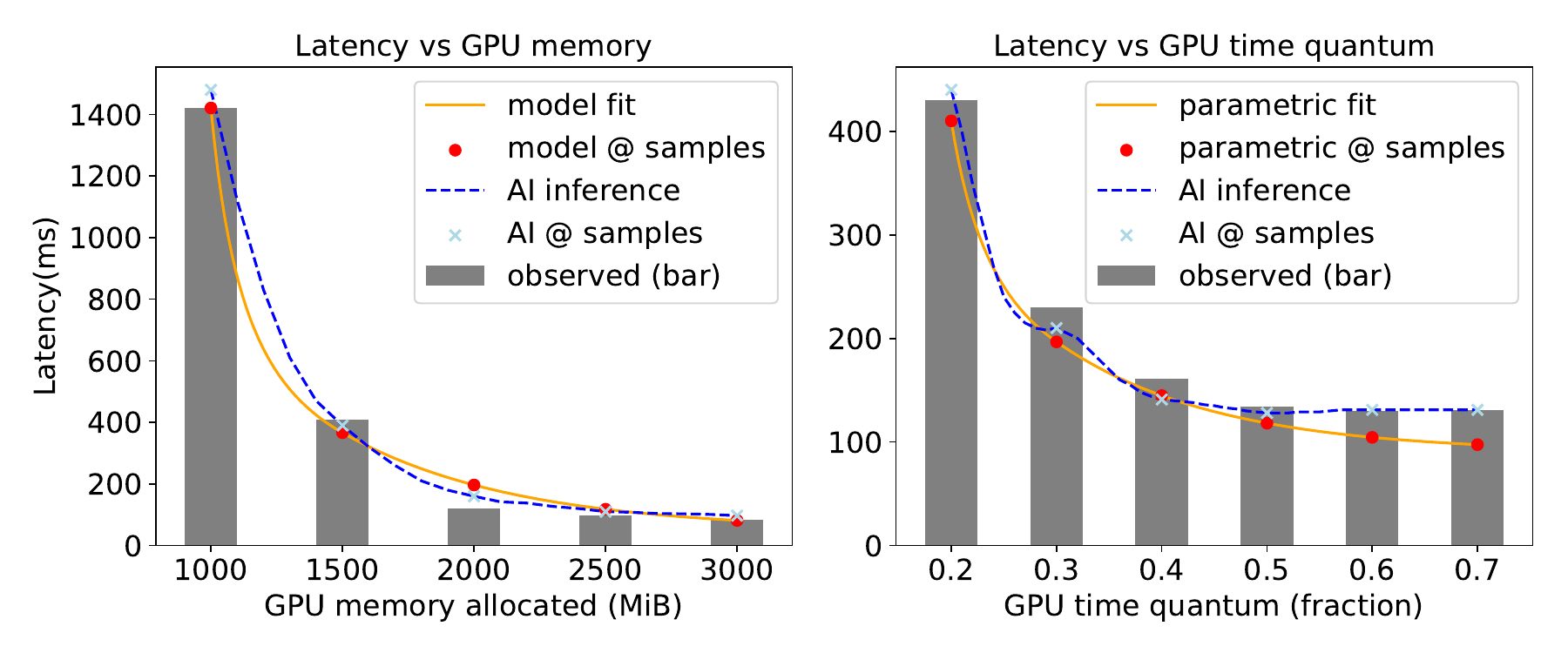}
  \caption{Predicted QoS versus ground truth for the parametric
    (orange) and deep-learning (blue) models.  Shaded regions indicate
    the 95\,\% confidence interval over five random train/val/test
    splits.}
  \label{fig:dl_compare}
\end{figure}

To decide which predictor underpins our allocation engine we benchmarked the
\textbf{domain-guided parametric model} against the \textbf{deep-learning (MLP)
  baseline} introduced in Section~\ref{sec:domain_parametric_qos}. Both were
trained on the same profiling dataset and evaluated on a disjoint test split.
Figure~\ref{fig:dl_compare} visualizes the fitted curves; the numerical errors
are summarized in Table~\ref{tab:model_comparison}.

\begin{table}[ht]
  \centering
  \caption{Mean-squared error (MSE) on the training and evaluation
    sets for latency (lower-is-better) and throughput
    (higher-is-better).}
  \label{tab:model_comparison}
  \begin{tabular}{lcc|cc}
    \toprule
                        & \multicolumn{2}{c|}{\textbf{Latency MSE}} & \multicolumn{2}{c}{\textbf{Throughput MSE}}                \\
    \cmidrule(lr){2-3} \cmidrule(lr){4-5}
    \textbf{Model}      & Train                                     & Eval                                        & Train & Eval \\
    \midrule
    Parametric guided   & 143                                       & 157                                         & 13.1  & 14.1 \\
    Deep learning (MLP) & 38.1                                      & 142                                         & 4.7   & 13.7 \\
    \bottomrule
  \end{tabular}
\end{table}

\smallskip
\paragraph{Findings}
The evaluation shows that although the MLP achieves the lowest training error,
its test-set MSE nearly matches that of the parametric model, signalling mild
over-fitting unless strong regularisation is applied. In contrast, the guided
model, governed by only a few physically interpretable parameters
\((\alpha_k,\beta_k)\), offers greater robustness and clarity, whereas the
MLP’s hundreds of weights obscure causal insight and impose checkpoint
management overhead. Moreover, the parametric form carries a negligible
operational footprint because it can be evaluated without a heavyweight
inference stack—an advantage on resource-constrained hypervisor hosts.

\smallskip
\paragraph{Decision}
Despite the MLP’s slightly better point accuracy, we adopt the domain-guided
parametric model because it resists over-fitting on unseen workloads, remains
transparent to operators and reviewers, and incurs virtually no storage or
compute cost. This choice aligns with our goal of delivering a lean,
explainable QoS predictor that generalizes across diverse VM configurations
without the burden of shipping additional AI artefacts.

\medskip
\section{Testbed}

All experiments were executed on an \emph{Intel Malibou Lake (MBL)} automotive
reference platform. The board integrates a 14-core heterogeneous CPU
configuration, consisting of up to 6 P-cores and 8 E-cores, paired with 64\,GB
of LPDDR5-6400 memory. Discrete graphics are provided by an \emph{Intel Arc
  A750 Automotive} GPU featuring SR-IOV, 28 Xe cores, and 16\,GB of on-board
GDDR6. System firmware follows the standard Intel boot flow, combining the
Firmware Support Package (FSP) with \emph{Slim
  Bootloader}\cite{slimbootloader}. A \texttt{ACRN} hypervisor\cite{li2019acrn}
boots first and hosts a Yocto-based \emph{Service
  OS}\cite{salvador2014embedded} that supplies device pass-through and
VM-orchestration utilities used throughout the study.

Tooling and image preparation are performed on a commodity development
workstation. While the build process does not require heavy computation, the
\texttt{acrn-configurator} mandates an \emph{Ubuntu 22.04} environment; any
modern desktop or laptop meeting this OS requirement suffices for build and
configuration tasks.

\medskip
\section{Implementation}

This section describes the detailed implementation process, covering VM
profiling, UI integration, resource allocation, and heuristic search tuning.

\smallskip
\subsection{VM Profiling}

The profiling stage is automated with script generation. After the user chooses
an OS image and high-level resource requirements (Application type, use of
peripherals like dGPU) for each VM, the system performs three steps:

\begin{enumerate}[label=\arabic*)]
  \item \textbf{Script synthesis}: generate an installation script that
        deploys the required monitoring utilities inside the guest.
  \item \textbf{Baseline run}: boot the VM with \emph{unconstrained}
        resources so the target workload can expose its natural demand
        envelope.
  \item \textbf{Trace collection}: execute the workload while recording
        time-stamped CPU, memory, GPU, I/O metrics.
\end{enumerate}

The resulting trace forms the profile vector~$\mathbf{p}$ and supplies the
ground truth for fitting both the domain-guided model and the deep-learning
baseline.

\smallskip
\subsection{UI Integration}

\begin{figure}[tbp]
  \centering
  \includegraphics[width=\linewidth]{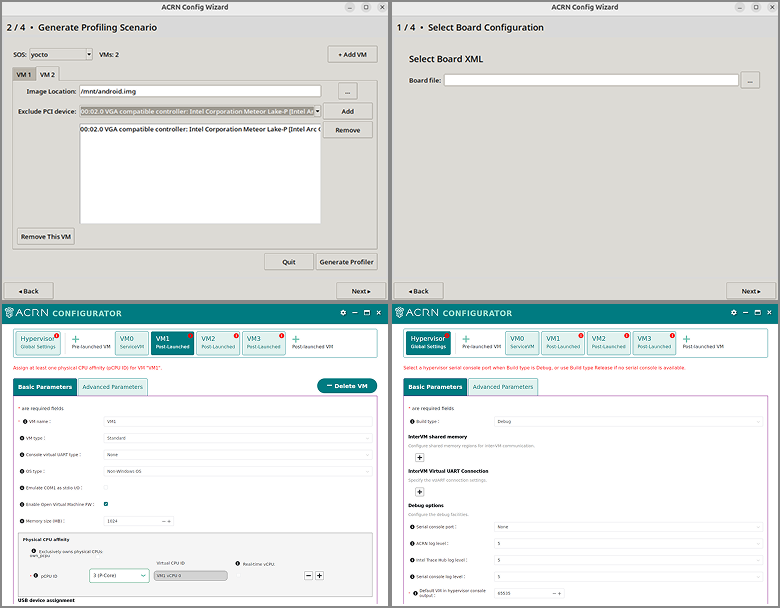}
  \caption{UI-integrated wizard tool (top) for the automated resource-allocation engine, and hypervisor scenario configuration tool (bottom).}
  \label{fig:ui_integration}
\end{figure}

As illustrated in Figure~\ref{fig:ui_integration}, the automated allocation
engine is integrated into a UI-based wizard tool. The process begins with the
user inspects the target board's hardware profile, so called as board
configuration. This configuration can be extracted using the ACRN hypervisor's
board configuration tool~\cite{li2019acrn}, which outputs the data in XML
format.

The optimization engine takes this XML file as input. The tool then guides the
integrator to specify high-level requirements for each virtual machine, such as
the operating system, workload class, peripheral devices, and other relevant
attributes. Based on these inputs, the UI generates and launches a tailored
monitoring shell script that collects time-series runtime metrics. Once
profiling is complete, the data is saved in CSV format and passed to the
resource allocation engine for optimization.

\smallskip
\subsection{Resource Allocation}

The time-series profiling data from each VM is transformed into a compact
vector format by extracting key features such as the maximum and median
utilization of CPU, memory, and GPU. This summarized data is then provided to
the optimization engine along with the system's hardware constraints.

For the resource allocation process, we define the optimization impact factor
in Equation~\ref{eq:qos_impact_factor}, which represents a weighted combination
of multiple QoS dimensions. The weights are configurable to reflect
workload-specific priorities—for example, gaming workloads may emphasize FPS
and latency (e.g., FPS: 8, latency: 2), while AI inference tasks may prioritize
throughput and latency (e.g., throughput: 7, latency: 3). The underlying QoS
model is vendor-defined and treated as a generalized function, though end-users
are afforded the flexibility to recalibrate these weights to better reflect
their operational characteristics.

Furthermore, the utilization parameter $\lambda_{\text{util}}$ in
Equation~\ref{eq:vm_opt_score} can be adapted and customized for each VM.
General-purpose VMs are typically assigned a lower value, indicating more
relaxed resource constraints, while specialized VMs may use higher values to
enforce tighter, efficiency-driven allocations. A recommended tuning range for
this parameter is between 0.01 and 0.1.

Given the QoS model and utilization parameters, the optimization engine
searches for the optimal resource allocation for each VM. Our implementation
employs a depth-first backtracking search algorithm, as outlined in
Algorithm~\ref{alg:backtracking}. The algorithm systematically explores all
possible allocation combinations for each VM, while pruning branches that
cannot yield a better score than the current best solution.

\begin{algorithm}[t]
  \label{alg:backtracking}
  \caption{Depth‑First Backtracking Allocator}
  \begin{algorithmic}[1]
    \Require Number of VMs $N$; host capacities $\mathbf{C}_{\text{tot}}$;
    candidate sets $\mathcal{C}_i$;
    utility $\mathrm{VMOptScore}$; bound $\mathrm{UPPER}$
    \Ensure  Optimal allocation $\{\mathbf{r}_i^\star\}_{i=1}^N$, best score $S^\star$
    \State $S^\star \gets -\infty$   \Comment{incumbent best score}
    \State $\mathbf{R}^\star \gets \varnothing$ \Comment{incumbent allocation}
    \State $\mathbf{R} \gets [\,]$              \Comment{stack of partial choices}
    \State $\mathbf{U} \gets \mathbf{0}$        \Comment{resources used so far}
    \smallskip
    \Procedure{Backtrack}{$(i,\,S)$}            \Comment{VM index $i$, current score $S$}
    \If{$i > N$}                              \Comment{all VMs assigned}
    \If{$S > S^\star$}
    \State $S^\star \gets S$;  $\mathbf{R}^\star \gets \mathbf{R}$
    \EndIf
    \State \textbf{return}
    \EndIf
    \State $\mathbf{R}_{\text{left}} \gets \mathbf{C}_{\text{tot}} - \mathbf{U}$
    \If{$S + \mathrm{UPPER}(i,\mathbf{R}_{\text{left}}) \le S^\star$}
    \State \textbf{return}                  \Comment{prune—cannot beat incumbent}
    \EndIf
    \ForAll{$\mathbf{r} \in \mathcal{C}_i$ \textbf{in heuristic order}}
    \If{$\mathbf{r} \le \mathbf{R}_{\text{left}}$ \textbf{and} $\mathbf{r}\in\mathcal{F}_i$}
    \State $\mathbf{R}.\text{push}(\mathbf{r})$;  $\mathbf{U} \gets \mathbf{U} + \mathbf{r}$
    \State \Call{Backtrack}{$i+1,\; S + \mathrm{VMOptScore}(i,\mathbf{r})$}
    \State $\mathbf{U} \gets \mathbf{U} - \mathbf{r}$;  $\mathbf{R}.\text{pop}()$
    \EndIf
    \EndFor
    \EndProcedure
    \smallskip
    \Statex
    \Comment{Optionally sort VMs/candidates first}
    \State \Call{Backtrack}{$1,\,0$}
    \State \Return $\{\mathbf{r}_i^\star\},\, S^\star$
  \end{algorithmic}
\end{algorithm}

Despite employing backtracking for allocation optimization, the computational
burden remains tractable due to the model's low complexity and the discretized
resource space. However, incorporating more granular resource search or
adopting computationally intensive QoS models (such as those based on machine
learning) may significantly increase the overall computational load.

The optimal allocation output is passed to the hypervisor configuration tools,
which generate one-shot VM launch scripts. These scripts are then transferred
to the target machine, where they are used to launch the VMs with the specified
resource allocations.

\subsection{Heuristic Search Tuning}
\label{subsec:heuristic_search_tuning}

Since any analytical model (whether domain-guided or neural) is ultimately an
estimate, the predicted "optimal" allocation may differ from the true QoS
optimum. Additionally, integrators may need to adjust the allocation to
accommodate unforeseen changes in workload behavior. In practice, this
refinement is often carried out through heuristic search, which involves
iteratively tweaking the allocation and measuring the resulting QoS. However,
this process is typically repetitive and time-consuming, and it is not
automated in our current implementation. Automating this refinement phase is
left as future work.

\smallskip
\subsection{Reduced Heuristic Search with Optimized Allocation}

For the measurement of the optimized allocation efficiency, we compared the
\emph{optimized allocation} from the analytic model against two baseline
strategies: \emph{equal split} and \emph{profile-proportional split}. The
\emph{equal split} strategy divides resources uniformly among VMs, while the
\emph{profile-proportional split} allocates resources in proportion to the
profiled demand of each VM. The goal is to determine how many trials are needed
to achieve the desired QoS criteria for each strategy, and how the optimized
allocation compares against these baselines.

\smallskip
\paragraph{Additional Search Steps}

\begin{itemize}
  \item \textbf{Incremental search.}
        At each trial we adjust one resource
        (CPU cores, memory, GPU slice) by its minimal quantum
        (e.g., $1$ core, $128$\,MiB memory).

  \item \textbf{Stopping rules.}
        Trial stops as soon as \emph{all} QoS metrics
        satisfy user-defined thresholds

  \item \textbf{Trial counter.}
        We count one trial per single-resource modification; this lets us
        compare search efficiency across strategies.

  \item \textbf{Baselines evaluated.}
        \begin{enumerate}[label=(\arabic*)]
          \item \emph{Equal split} - divide resources uniformly among VMs.
          \item \emph{Profile-proportional split} - allocate in proportion to
                profiled demand.
          \item \emph{Optimized split} - start from the analytic
                optimum and refine via the heuristic search above.
        \end{enumerate}
\end{itemize}

\begin{figure}[tbp]
  \centering
  \includegraphics[width=\linewidth]{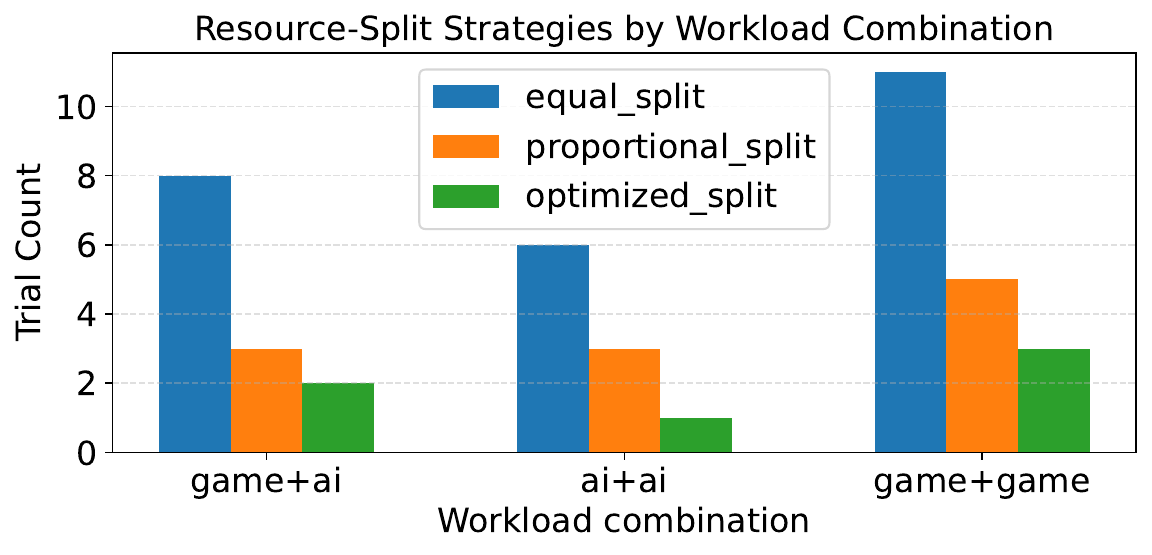}
  \caption{Heuristic search trial count for different allocation strategies.}
  \label{fig:search_trials}
\end{figure}

Figure~\ref{fig:search_trials} shows that the \emph{optimized + refine}
strategy achieves the QoS criteria while requiring \emph{fewer} trials than
either equal or proportional allocation, demonstrating the value of combining
modeling with targeted on-line exploration.

\smallskip
\subsection{Additional Benefits of the Guided Resource Allocation}

While this paper primarily focuses on the performance impact of optimized
resource allocation, one of the key benefits of the proposed process is its
ability to reduce the risk of resource misallocation. For example, a customer
might assign a mix of P-cores and E-cores to an RTOS, which is not recommended
by the vendor. In addition, as discussed in
Section~\ref{sec:problem_statement}, the recommended size of PF memory depends
on the display resolution. If allocated improperly, this can lead to system
inefficiencies. Similarly, parameters such as time quantum and preemptive
timeout for dGPU SR-IOV have recommended ranges and are prone to
misconfiguration. The proposed guided allocation process helps prevent such
misconfigurations, which are common in manually configured environments.

\medskip
\section{Discussion}

\smallskip
\subsection{Field Application Report}

\begin{figure}[htbp]
  \centering
  \includegraphics[width=\linewidth]{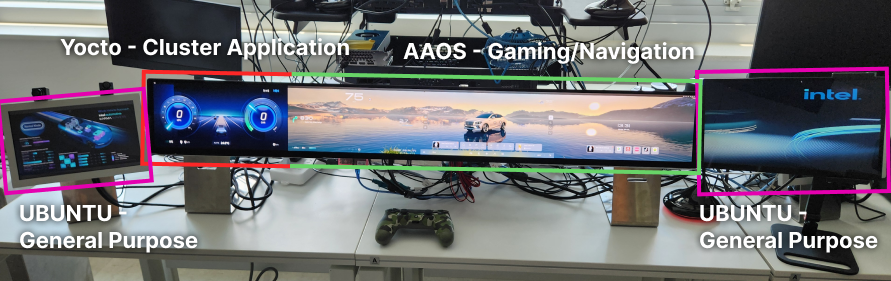}
  \caption{Field application of the resource allocation process, demonstrating VM partitioning with optimized resource allocation. Yocto Linux is used for the cluster application, AAOS for high-perf gaming and navigation, and Ubuntu 22.04 for the general-purpose VM.}
  \label{fig:field_application}
\end{figure}

The proposed resource allocation process has been successfully applied to
real-world scenarios involving VM partitioning for heterogeneous workloads. As
illustrated in Figure~\ref{fig:field_application}, our deployment includes a
Unity engine application running on AAOS and a visualized cluster running on
Yocto Linux, both sharing system resources. In addition, Ubuntu 22.04 is used
as a general-purpose VM, sharing the dGPU with AAOS via SR-IOV. This approach
not only reduces the risk of systemic issues in our implementation, but also
minimizes the chances for customers to encounter performance degradation or
instability caused by poor resource allocation, which is an issue often seen in
manually configured environments. Although the current workflow still requires
VM profiling and heuristic refinement after the initial optimization, these
steps remain a burden in the development cycle and highlight the need for
further automation to streamline integration and reduce human error.

\smallskip
\subsection{Future Work}

The following points remain open for future work:
\begin{itemize}
  \item \textbf{Automating heuristic tuning.} Beyond the current manually guided search, we plan to automate the heuristic-search phase itself, similar to hyperparameter grid search in deep learning. This would allow additional tuning to be carried out autonomously, further accelerating convergence toward near-optimal allocations.
  \item \textbf{Learning-based decision engines.} By integrating LLM assistants\cite{lee2024llm,liu2024resource}, multi-armed bandits (MAB)\cite{galli2023playing,zuo2021combinatorial}, or reinforcement-learning agents, the system can automatically suggest or enact mitigation and allocation strategies as workload patterns evolve.
  \item \textbf{Dynamic, adaptive allocation.} We aim to extend the framework so that it continuously re-evaluates runtime telemetry and adjusts resource assignments without redeployment, maintaining optimal performance under changing conditions\cite{xiao2012dynamic}.
  \item \textbf{Real-time-aware policies.} Finally, supporting strict real-time constraints, such as latency budgets and jitter bounds, remains largely unexplored in the current prototype and can be a focus of subsequent research\cite{zou2025survey}.
\end{itemize}

\medskip
\section{Acknowledgments}
We extend our sincere thanks to \emph{Intel}, particularly the
\emph{Automotive} team, for their support and valuable contributions to this
work.

\bibliographystyle{IEEEtran}
\bibliography{refs}

\begin{thebibliography}{10}
\providecommand{\url}[1]{#1}
\csname url@samestyle\endcsname
\providecommand{\newblock}{\relax}
\providecommand{\bibinfo}[2]{#2}
\providecommand{\BIBentrySTDinterwordspacing}{\spaceskip=0pt\relax}
\providecommand{\BIBentryALTinterwordstretchfactor}{4}
\providecommand{\BIBentryALTinterwordspacing}{\spaceskip=\fontdimen2\font plus
\BIBentryALTinterwordstretchfactor\fontdimen3\font minus \fontdimen4\font\relax}
\providecommand{\BIBforeignlanguage}[2]{{%
\expandafter\ifx\csname l@#1\endcsname\relax
\typeout{** WARNING: IEEEtran.bst: No hyphenation pattern has been}%
\typeout{** loaded for the language `#1'. Using the pattern for}%
\typeout{** the default language instead.}%
\else
\language=\csname l@#1\endcsname
\fi
#2}}
\providecommand{\BIBdecl}{\relax}
\BIBdecl

\bibitem{liu2022impact}
Z.~Liu, W.~Zhang, and F.~Zhao, ``Impact, challenges and prospect of software-defined vehicles,'' \emph{Automotive Innovation}, vol.~5, no.~2, pp. 180--194, 2022.

\bibitem{jiacheng2016software}
C.~Jiacheng, Z.~Haibo, Z.~Ning, Y.~Peng, G.~Lin, and S.~X. Sherman, ``Software defined internet of vehicles: architecture, challenges and solutions,'' \emph{Journal of communications and information networks}, vol.~1, no.~1, pp. 14--26, 2016.

\bibitem{bhatia2019software}
J.~Bhatia, Y.~Modi, S.~Tanwar, and M.~Bhavsar, ``Software defined vehicular networks: A comprehensive review,'' \emph{International Journal of Communication Systems}, vol.~32, no.~12, p. e4005, 2019.

\bibitem{papazoglou2007service}
M.~P. Papazoglou and W.-J. Van Den~Heuvel, ``Service oriented architectures: approaches, technologies and research issues,'' \emph{The VLDB journal}, vol.~16, no.~3, pp. 389--415, 2007.

\bibitem{valipour2009brief}
M.~H. Valipour, B.~AmirZafari, K.~N. Maleki, and N.~Daneshpour, ``A brief survey of software architecture concepts and service oriented architecture,'' in \emph{2009 2nd IEEE International Conference on Computer Science and Information Technology}.\hskip 1em plus 0.5em minus 0.4em\relax IEEE, 2009, pp. 34--38.

\bibitem{hustad2021creating}
E.~Hustad and D.~H. Olsen, ``Creating a sustainable digital infrastructure: The role of service-oriented architecture,'' \emph{Procedia Computer Science}, vol. 181, pp. 597--604, 2021.

\bibitem{mauser2025centralization}
L.~Mauser and S.~Wagner, ``Centralization potential of automotive e/e architectures,'' \emph{Journal of Systems and Software}, vol. 219, p. 112220, 2025.

\bibitem{bandur2021making}
V.~Bandur, G.~Selim, V.~Pantelic, and M.~Lawford, ``Making the case for centralized automotive e/e architectures,'' \emph{IEEE Transactions on Vehicular Technology}, vol.~70, no.~2, pp. 1230--1245, 2021.

\bibitem{kampmann2022optimization}
A.~Kampmann, M.~L{\"u}er, S.~Kowalewski, and B.~Alrifaee, ``Optimization-based resource allocation for an automotive service-oriented software architecture,'' in \emph{2022 IEEE Intelligent Vehicles Symposium (IV)}.\hskip 1em plus 0.5em minus 0.4em\relax IEEE, 2022, pp. 678--687.

\bibitem{reinhardt2014embedded}
D.~Reinhardt and G.~Morgan, ``An embedded hypervisor for safety-relevant automotive e/e-systems,'' in \emph{Proceedings of the 9th IEEE International Symposium on Industrial Embedded Systems (SIES 2014)}.\hskip 1em plus 0.5em minus 0.4em\relax IEEE, 2014, pp. 189--198.

\bibitem{jiang2024towards}
L.~Jiang, F.~Zhang, and J.~Ming, ``Towards intelligent automobile cockpit via a new container architecture,'' in \emph{21st USENIX Symposium on Networked Systems Design and Implementation (NSDI 24)}, 2024, pp. 205--219.

\bibitem{karthik2018hypervisor}
S.~Karthik, K.~Ramanan, N.~Devshatwar, S.~Paul, V.~Mahaveer, S.~Zhao, M.~Vishwanathan, and C.~Matad, ``Hypervisor based approach for integrated cockpit solutions,'' in \emph{2018 IEEE 8th international conference on consumer electronics-berlin (ICCE-Berlin)}.\hskip 1em plus 0.5em minus 0.4em\relax IEEE, 2018, pp. 1--6.

\bibitem{li2019acrn}
H.~Li, X.~Xu, J.~Ren, and Y.~Dong, ``Acrn: a big little hypervisor for iot development,'' in \emph{Proceedings of the 15th ACM SIGPLAN/SIGOPS International Conference on Virtual Execution Environments}, 2019, pp. 31--44.

\bibitem{zhang2022innovation}
Y.~Zhang, ``Innovation dynamics between original equipment manufacturers (oems) and tier-1 suppliers in the automotive industry,'' Ph.D. dissertation, Massachusetts Institute of Technology, 2022.

\bibitem{volpato2004oem}
G.~Volpato, ``The oem-fts relationship in automotive industry,'' \emph{International Journal of Automotive Technology and Management}, vol.~4, no. 2-3, pp. 166--197, 2004.

\bibitem{trovao2019trends}
J.~P. Trovao, ``Trends in automotive electronics [automotive electronics],'' \emph{IEEE Vehicular Technology Magazine}, vol.~14, no.~4, pp. 100--109, 2019.

\bibitem{BlackBerryQNX_HypervisorAutomotive2023}
\BIBentryALTinterwordspacing
B.~QNX, ``Leveraging {QNX Hypervisor} for automotive {E/E} architecture consolidation,'' BlackBerry QNX, Tech. Rep., 2023, white Paper. [Online]. Available: \url{blackberry.qnx.com/content/dam/qnx/products/hypervisor/hypervisorAutomotive-ProductBrief.pdf}
\BIBentrySTDinterwordspacing

\bibitem{baryshnikov2016jailhouse}
M.~Baryshnikov, ``Jailhouse hypervisor,'' {B.S.} thesis, {\v{C}}esk{\'e} vysok{\'e} u{\v{c}}en{\'\i} technick{\'e} v Praze. Vypo{\v{c}}etn{\'\i} a informa{\v{c}}n{\'\i} centrum., 2016.

\bibitem{lozano2023comprehensive}
S.~Lozano, T.~Lugo, and J.~Carretero, ``A comprehensive survey on the use of hypervisors in safety-critical systems,'' \emph{IEEE Access}, vol.~11, pp. 36\,244--36\,263, 2023.

\bibitem{macario2009vehicle}
G.~Macario, M.~Torchiano, and M.~Violante, ``An in-vehicle infotainment software architecture based on google android,'' in \emph{2009 IEEE International Symposium on Industrial Embedded Systems}.\hskip 1em plus 0.5em minus 0.4em\relax IEEE, 2009, pp. 257--260.

\bibitem{jeong2023infotainment}
S.~Jeong, M.~Ryu, H.~Kang, and H.~K. Kim, ``Infotainment system matters: Understanding the impact and implications of in-vehicle infotainment system hacking with automotive grade linux,'' in \emph{Proceedings of the Thirteenth ACM Conference on Data and Application Security and Privacy}, 2023, pp. 201--212.

\bibitem{sohn2024strategy}
K.~Sohn, I.~Choi, S.~Kim, J.~Lee, J.~Lee, and J.~Kim, ``A strategy to maximize the utilization of ai neural processors on an automotive computing platform,'' in \emph{2024 IEEE International Conference on Consumer Electronics (ICCE)}.\hskip 1em plus 0.5em minus 0.4em\relax IEEE, 2024, pp. 1--4.

\bibitem{dong2012high}
Y.~Dong, X.~Yang, J.~Li, G.~Liao, K.~Tian, and H.~Guan, ``High performance network virtualization with sr-iov,'' \emph{Journal of Parallel and Distributed Computing}, vol.~72, no.~11, pp. 1471--1480, 2012.

\bibitem{younge2015supporting}
A.~J. Younge, J.~P. Walters, S.~P. Crago, and G.~C. Fox, ``Supporting high performance molecular dynamics in virtualized clusters using iommu, sr-iov, and gpudirect,'' \emph{ACM SIGPLAN Notices}, vol.~50, no.~7, pp. 31--38, 2015.

\bibitem{boyd2004convex}
S.~P. Boyd and L.~Vandenberghe, \emph{Convex optimization}.\hskip 1em plus 0.5em minus 0.4em\relax Cambridge university press, 2004.

\bibitem{bertsekas2009convex}
D.~Bertsekas, \emph{Convex optimization theory}.\hskip 1em plus 0.5em minus 0.4em\relax Athena Scientific, 2009, vol.~1.

\bibitem{bubeck2015convex}
S.~Bubeck \emph{et~al.}, ``Convex optimization: Algorithms and complexity,'' \emph{Foundations and Trends{\textregistered} in Machine Learning}, vol.~8, no. 3-4, pp. 231--357, 2015.

\bibitem{lai1985asymptotically}
T.~L. Lai and H.~Robbins, ``Asymptotically efficient adaptive allocation rules,'' \emph{Advances in applied mathematics}, vol.~6, no.~1, pp. 4--22, 1985.

\bibitem{auer2002finite}
P.~Auer, N.~Cesa-Bianchi, and P.~Fischer, ``Finite-time analysis of the multiarmed bandit problem,'' \emph{Machine learning}, vol.~47, no.~2, pp. 235--256, 2002.

\bibitem{vermorel2005multi}
J.~Vermorel and M.~Mohri, ``Multi-armed bandit algorithms and empirical evaluation,'' in \emph{European conference on machine learning}.\hskip 1em plus 0.5em minus 0.4em\relax Springer, 2005, pp. 437--448.

\bibitem{galli2023playing}
A.~Galli, V.~Moscato, S.~P. Romano, and G.~Sperl{\'\i}, ``Playing with a multi armed bandit to optimize resource allocation in satellite-enabled 5g networks,'' \emph{IEEE Transactions on Network and Service Management}, vol.~21, no.~1, pp. 341--354, 2023.

\bibitem{wolsey1999integer}
L.~A. Wolsey and G.~L. Nemhauser, \emph{Integer and combinatorial optimization}.\hskip 1em plus 0.5em minus 0.4em\relax John Wiley \& Sons, 1999.

\bibitem{jain2001algorithms}
V.~Jain and I.~E. Grossmann, ``Algorithms for hybrid milp/cp models for a class of optimization problems,'' \emph{INFORMS Journal on computing}, vol.~13, no.~4, pp. 258--276, 2001.

\bibitem{osman1996meta}
I.~H. Osman and J.~P. Kelly, ``Meta-heuristics: an overview,'' \emph{Meta-heuristics: Theory and applications}, pp. 1--21, 1996.

\bibitem{kirkpatrick1983sa}
S.~Kirkpatrick, C.~D. Gelatt, and M.~P. Vecchi, ``Optimization by simulated annealing,'' \emph{Science}, vol. 220, pp. 671--680, 1983.

\bibitem{sun2020optimal}
W.~Sun, Y.~Wang, and S.~Li, ``An optimal resource allocation scheme for virtual machine placement of deploying enterprise applications into the cloud,'' \emph{AIMS Mathematics}, vol.~5, no.~4, pp. 3966--3989, 2020.

\bibitem{dubey2023resource}
K.~Dubey, S.~Sharma, and M.~Kumar, ``Resource optimization based virtual machine allocation technique in cloud computing domain,'' in \emph{2023 14th international conference on computing communication and networking technologies (ICCCNT)}.\hskip 1em plus 0.5em minus 0.4em\relax IEEE, 2023, pp. 1--7.

\bibitem{kabir2023virtualization}
M.~R. Kabir and S.~Ray, ``Virtualization for automotive safety and security exploration,'' in \emph{2023 IEEE 16th Dallas circuits and systems conference (DCAS)}.\hskip 1em plus 0.5em minus 0.4em\relax IEEE, 2023, pp. 1--4.

\bibitem{rao2009vconf}
J.~Rao, X.~Bu, C.-Z. Xu, L.~Wang, and G.~Yin, ``Vconf: a reinforcement learning approach to virtual machines auto-configuration,'' in \emph{Proceedings of the 6th international conference on Autonomic computing}, 2009, pp. 137--146.

\bibitem{hummaida2022scalable}
A.~R. Hummaida, N.~W. Paton, and R.~Sakellariou, ``Scalable virtual machine migration using reinforcement learning,'' \emph{Journal of Grid Computing}, vol.~20, no.~2, p.~15, 2022.

\bibitem{ma2022real}
X.~Ma, H.~Xu, H.~Gao, M.~Bian, and W.~Hussain, ``Real-time virtual machine scheduling in industry iot network: A reinforcement learning method,'' \emph{IEEE Transactions on Industrial Informatics}, vol.~19, no.~2, pp. 2129--2139, 2022.

\bibitem{shah2020multiagent}
H.~A. Shah and L.~Zhao, ``Multiagent deep-reinforcement-learning-based virtual resource allocation through network function virtualization in internet of things,'' \emph{IEEE Internet of Things Journal}, vol.~8, no.~5, pp. 3410--3421, 2020.

\bibitem{khan2022workload}
T.~Khan, W.~Tian, S.~Ilager, and R.~Buyya, ``Workload forecasting and energy state estimation in cloud data centres: Ml-centric approach,'' \emph{Future Generation Computer Systems}, vol. 128, pp. 320--332, 2022.

\bibitem{gong2024dynamic}
Y.~Gong, J.~Huang, B.~Liu, J.~Xu, B.~Wu, and Y.~Zhang, ``Dynamic resource allocation for virtual machine migration optimization using machine learning,'' \emph{arXiv preprint arXiv:2403.13619}, 2024.

\bibitem{vhatkar2024improved}
K.~Vhatkar, A.~B. Kathole, A.~P. Kshirsagar, and J.~Katti, ``Improved optimization algorithm for resource management in cloud applications with performance monitor of vm provisioning, placement and recycling,'' \emph{Journal of High Speed Networks}, vol.~30, no.~4, pp. 583--606, 2024.

\bibitem{nikov2015evaluation}
K.~Nikov, J.~L. Nunez-Yanez, and M.~Horsnell, ``Evaluation of hybrid run-time power models for the arm big. little architecture,'' in \emph{2015 IEEE 13th International Conference on Embedded and Ubiquitous Computing}.\hskip 1em plus 0.5em minus 0.4em\relax IEEE, 2015, pp. 205--210.

\bibitem{burgio2017software}
P.~Burgio, M.~Bertogna, N.~Capodieci, R.~Cavicchioli, M.~Sojka, P.~Houdek, A.~Marongiu, P.~Gai, C.~Scordino, and B.~Morelli, ``A software stack for next-generation automotive systems on many-core heterogeneous platforms,'' \emph{Microprocessors and Microsystems}, vol.~52, pp. 299--311, 2017.

\bibitem{wang2023efficient}
Y.~Wang, B.~Luo, and Y.~Shen, ``Efficient memory overcommitment for $\{$I/O$\}$ passthrough enabled $\{$VMs$\}$ via fine-grained page meta-data management,'' in \emph{2023 USENIX Annual Technical Conference (USENIX ATC 23)}, 2023, pp. 769--783.

\bibitem{sawamura2015evaluating}
R.~Sawamura, C.~Boeres, and V.~E. Rebello, ``Evaluating the impact of memory allocation and swap for vertical memory elasticity in vms,'' in \emph{2015 27th International Symposium on Computer Architecture and High Performance Computing (SBAC-PAD)}.\hskip 1em plus 0.5em minus 0.4em\relax IEEE, 2015, pp. 186--193.

\bibitem{shin2024beyond}
S.~Shin, S.~J. Chae, S.~Lee, and J.~K. Kim, ``Beyond homogeneity: Assessing the validity of the michaelis--menten rate law in spatially heterogeneous environments,'' \emph{PLOS Computational Biology}, vol.~20, no.~6, p. e1012205, 2024.

\bibitem{sodhro2019artificial}
A.~H. Sodhro, Z.~Luo, G.~H. Sodhro, M.~Muzamal, J.~J. Rodrigues, and V.~H.~C. De~Albuquerque, ``Artificial intelligence based qos optimization for multimedia communication in iov systems,'' \emph{Future Generation Computer Systems}, vol.~95, pp. 667--680, 2019.

\bibitem{mao2020ai}
B.~Mao, Y.~Kawamoto, and N.~Kato, ``Ai-based joint optimization of qos and security for 6g energy harvesting internet of things,'' \emph{IEEE Internet of Things Journal}, vol.~7, no.~8, pp. 7032--7042, 2020.

\bibitem{slimbootloader}
{Intel Corporation}, ``Slim bootloader,'' \url{https://slimbootloader.github.io/}, 2024, accessed: 2025-07-27.

\bibitem{salvador2014embedded}
O.~Salvador and D.~Angolini, \emph{Embedded Linux Development with Yocto Project}.\hskip 1em plus 0.5em minus 0.4em\relax Packt Publishing, 2014.

\bibitem{lee2024llm}
W.~Lee and J.~Park, ``Llm-empowered resource allocation in wireless communications systems,'' \emph{arXiv preprint arXiv:2408.02944}, 2024.

\bibitem{liu2024resource}
C.~Liu and J.~Zhao, ``Resource allocation for stable llm training in mobile edge computing,'' in \emph{Proceedings of the Twenty-fifth International Symposium on Theory, Algorithmic Foundations, and Protocol Design for Mobile Networks and Mobile Computing}, 2024, pp. 81--90.

\bibitem{zuo2021combinatorial}
J.~Zuo and C.~Joe-Wong, ``Combinatorial multi-armed bandits for resource allocation,'' in \emph{2021 55th Annual Conference on Information Sciences and Systems (CISS)}.\hskip 1em plus 0.5em minus 0.4em\relax IEEE, 2021, pp. 1--4.

\bibitem{xiao2012dynamic}
Z.~Xiao, W.~Song, and Q.~Chen, ``Dynamic resource allocation using virtual machines for cloud computing environment,'' \emph{IEEE transactions on parallel and distributed systems}, vol.~24, no.~6, pp. 1107--1117, 2012.

\bibitem{zou2025survey}
A.~Zou, Y.~Xu, Y.~Ni, J.~Chen, Y.~Ma, J.~Li, C.~Gill, X.~Zhang, and Y.~Jin, ``A survey of real-time scheduling on accelerator-based heterogeneous architecture for time critical applications,'' \emph{arXiv preprint arXiv:2505.11970}, 2025.

\end{thebibliography}

\vspace{12pt}


\end{document}